\documentclass[preprint2,longabstract]{aastex}

\renewcommand{\b}[1]{\mbox{\boldmath $#1$}}

\def\cal#1{{\cal #1}}

\def\m@th{\mathsurround=0pt}
\def\n@space{\nulldelimiterspace=0pt \m@th}
\def\biggg#1{{\mbox{$\left#1\vbox to 20.5pt{}\right.\n@space$}}}

\def\beginenum{\begin{enumerate}}
\def\endenum{\end{enumerate}}
\def\bitem{\begin{itemize}}
\def\eitem{\end{itemize}}
\def\bray{\begin{array}}
\def\eray{\end{array}}
\def\begindoc{\begin{document}}
\def\enddoc{\end{document}}
\def\bq{\begin{equation}}
\def\eq{\end{equation}}
\def\bqy{\begin{eqnarray}}
\def\eqy{\end{eqnarray}}
\def\bqyn{\begin{eqnarray*}}
\def\eqyn{\end{eqnarray*}}
\def\bc{\begin{center}}
\def\ec{\end{center}}
\def\bfll{\begin{flushleft}}
\def\efll{\end{flushleft}}
\def\bflr{\begin{flushright}}
\def\eflr{\end{flushright}}
\newcommand{\Avec}{\mbox{\boldmath $A$}}
\newcommand{\Bvec}{\mbox{\boldmath $B$}}
\newcommand{\Evec}{\mbox{\boldmath $E$}}
\newcommand{\Fvec}{\mbox{\boldmath $F$}}
\newcommand{\Gvec}{\mbox{\boldmath $G$}}
\newcommand{\Rvec}{\mbox{\boldmath $R$}}
\newcommand{\Uvec}{\mbox{\boldmath $U$}}
\newcommand{\Vvec}{\mbox{\boldmath $V$}}
\newcommand{\evec}{\mbox{\boldmath $e$}}
\newcommand{\jvec}{\mbox{\boldmath $j$}}
\newcommand{\kvec}{\mbox{\boldmath $k$}}
\newcommand{\nvec}{\mbox{\boldmath $n$}}
\newcommand{\uvec}{\mbox{\boldmath $u$}}
\newcommand{\vvec}{\mbox{\boldmath $v$}}
\newcommand{\wvec}{\mbox{\boldmath $w$}}
\newcommand{\xvec}{\mbox{\boldmath $x$}}
\newcommand{\omegavec}{\mbox{\boldmath $\omega$}}
\newcommand{\Omegavec}{\mbox{\boldmath $\Omega$}}

\usepackage{spr-astr-addons}    

\usepackage{color}

\RequirePackage{color}
\def\imagei{\centerline{\color[gray]{.75}\rule{\hsize}{4pc}}}%
\def\imageii{\centerline{\color[gray]{.75}\rule{4pc}{4pc}}}%

\newcommand{\vdag}{(v)^\dagger}
\newcommand{\emaila}{authors@email.com}

\begin{document}
%

\title{Catastrophic Formation of Macro-Scale Flow and Magnetic Fields
in the Relativistic Gas of Binary Systems}

\shorttitle{<Catastrophic Formation of Macro-field in Binary Systems>}
\shortauthors{< Dadiani, Saralidze, Shatashvili \& Mahajan>}

\author{ E. Saralidze\altaffilmark{1,4}}
\altaffiltext{1}{Department of Physics, Faculty of Exact \&
Natural Sciences, Javakhishvili Tbilisi State University,
Tbilisi 0179, Georgia}
\altaffiltext{4}{Department of Physics, College of Science,
North Carolina State University, 401 Stinson Drive, Raleigh, NC 27695-8202, USA}
\author{ N.L. Shatashvili\altaffilmark{1,2}}
\altaffiltext{1}{Department of Physics, Faculty of Exact \&
Natural Sciences, Javakhishvili Tbilisi State University,
Tbilisi 0179, Georgia}
\altaffiltext{2}{Andronikashvili Institute
of Physics, TSU, Tbilisi 0177, Georgia}
\author{ S.M. Mahajan\altaffilmark{3}}
\altaffiltext{3}{Institute for Fusion Studies, The University of
Texas at Austin, Austin,Tx 78712, USA} \and
\author{ E. Dadiani\altaffilmark{5}}
\altaffiltext{5}{McWilliams Center for Cosmology, Department of Physics,
Carnegie Mellon University, Pittsburgh, PA 15213, USA}

\begin{abstract}
It is shown that a simple quasi--equilibrium analysis of a multi-component
plasma can be harnessed to explain catastrophic energy transformations
in astrophysical objects. We limit ourselves to the particular class of
binary systems for which the  typical plasma consists of one classical
ion  component, and two relativistic electron components -- the bulk degenerate
electron gas with a small contamination of hot electrons. We derive,
analytically, the conditions conducive to such a catastrophic change.
The pathway to such sudden changes is created by the slow changes in
the initial parameters so that the governing equilibrium state can no
longer be sustained and the system must find a new equilibrium that
could have  vastly different energy mix-- of  thermal, flow--kinetic
and magnetic energies.  In one such scenario, macro--scale flow kinetic,
and magnetic energies abound in the final state. For the given
multi--component plasma, we show that the flow (strongly
Super--Alfv\'enic) kinetic energy is mostly carried by the small
hot electron component. Under specific conditions, it
is possible to generate strong macro--scale magnetic (velocity) field
when all of the flow (magnetic) field energy is converted to the magnetic
(velocity) field energy at the catastrophe. The analysis is applied to
explain various observed characteristics of white dwarf (WD) systems, in
particular, of the magnetic and dense/degenerate type.
\end{abstract}


\maketitle


\maketitle

\section{Introduction}

\subsection{Accreting White Dwarf Systems}

Most astrophysical ``Objects'' may be classed as multi-temperature
multi-species systems. A compact object like a White Dwarf (WD), for instance,
has a highly degenerate plasma co-existing with a classical hot
accreting flow. WDs comprise up to $90\%$ of the end state of stellar evolution
\citep{winget,Compact-WD,kulebi,kepler}, \citep{Shapiro}. Accreting
WDs (AWD), in addition, feature global magnetic structures with
typical field strengths (B=$1 -– 1000$)\,MG \\
\citep{White,hmfwd};\\
\citep{Kawka}.

Isolated WDs, however, can have much stronger fields and may be
separated into two categories: 1) the High Field Magnetic WDs (HFMWD) with
$B>10^6$G which may have binary origin according to recent studies
(see e.g., \citep{Garcia} and references therein), and 2) comparatively
lower field systems with $B<10^5$G \citep{DAWD,Kawka,Ferrario-3}.

Binary evolution of such AWDs (with rapid differential rotation)
is used to explain the Type Ia Supernovae high mass progenitors
(see e.g., \citep{hachisu,yoon} and references therein).
In addition to their own accretion, AWDs are often
surrounded by an accreting gas of a companion star / disk
\citep{Begelman,mukai}.

The representative accreting white dwarf binaries (AWBs) 
constitute star types called Cataclysmic Variables (CVs) that fall
into two classes:

1) Nonmagnetic with weak or nonexistent magnetic fields
($< 0.01MG$)); such CVs display eruptive behavior
\citep{Warner,DAWD,Balman,mukai},

2) the Magnetic CVs (MCVs) among which $25\%$ are very
magnetic \citep{Ferrario-2,Balman,Mouchet}; If the two stars merge
the end product is a single HFMWD that may later evolve into an
MCV \citep{Ferrario-4,Ferrario-3}.

AWBs are important laboratories 
[see e.g., \\
\cite{AD_modeling,kafka}; \\
\cite{Diaz}] for large--scale outflow physics
which, along with the magnetic field, plays an important role
in stellar evolution.

The first observational evidence for the presence of active,
localized magnetic structures in WDs was discussed in 
Valyavin et al (2011);
specifically, it was reported that the photosphere of WD1953--011
was endowed with a two--component magnetic field geometry made
up of  a weak, large--scale component, and a strong, localized
component (magnetic ``spot'') similar to the Sun.

One can bring more examples from existing rich
phenomenology on AWD systems but we will not present them here;
the reader may consult \\
Kotorashvili \& Shatashvili (2022) and \\
references therein.

In order to explicate/explore such richness we need a
unified theoretical framework that deals simultaneously with flows
and fields in a multi-component plasma. In this paper, we will, mostly,
invoke a framework  that was first developed to deal with both quiescent
and explosive phenomena in the solar atmosphere. Starting from the
formulation of   the general global dynamics that may operate in a given
atmospheric region \citep{mmns-1}, a specific model for catastrophic
energy transformations, that could take place in  the solar atmosphere
filled with a two component plasma, was developed in \citep{osym,osym2};
it was later extended to  several other astrophysical
settings \citep{kagan,AccrVort}; \\
{\citep{BS-flow,gondal}.

\bigskip

\subsection{Towards the quasi--equilibrium approach for
the energy transformations in AWD systems}

The main theme of the model developed in \\
\citep{osym,osym2} lies in what may be called a quasi--equilibrium
approach to predicting catastrophic energy transformations; it does not
actually deal with the dynamics of the catastrophe itself but shows
how slow changes in the parameters that label an equilibrium state could
drive the system to a stage where the original equilibrium can no longer
be sustained. Perforce, the system must be either ``destroyed'' or find
a new equilibrium; in either case the energy mix of the system could
be drastically changed.

In the present study, we will apply this well-tested methodology to
understand the explosive events and mass outflows for the AWD
systems. The plasma physics, for this case, is a little more complicated
than for the Solar Atmosphere case because of two
reasons: 1) we have an additional (lower density) hot electron
component, and 2) the high density bulk electrons are degenerate.  It is worth
mentioning that we have some prior experience dealing with multi-component
relativistic plasmas with a degenerate component \citep{BSM_deg};\\
\citep{SMB_multi,BS-flow,SMB_2T,RD_deg}; \\
\citep{KS_2TRD}.

The starting point for exploring the quasi--equilibrium approach to
catastrophic events is, naturally, the existence of a well defined
equilibrium that can be appropriately labelled by identifiable physical
parameters. The equilibria we deal with are the so called Multiple--Beltrami
relaxed states which are obtained by minimizing the total energy of
the plasma (thermal, kinetic, and electromagnetic) subject to the
so called helicity constraints. Each plasma species has its own
characteristic helicity invariant and these invariants are the
appropriate labels for an equilibrium.

The reader is referred to  considerable literature on Multi- Beltrami
relaxed states  \citep{MY,YM,ymois};\\
\citep{mmns-1,osym};\\
\citep{osym2}; (later, \citep{iqbal};\\
\citep{SMB_multi,SMB_2T}).
These relaxed states (force free in a generalized sense), derived by
the constrained minimization of total energy, are defined by a set
of simultaneous Beltrami conditions each signifying the alignment of
a species' velocity and its generalized vorticity. This class of
states will form the basis on which this study is constructed.

In particular, we will investigate,  in detail, the evolution of
relaxed states accessible to the three component plasma consisting of:
1) a mobile classical ion component, 2) and two relativistic electron species
-- the bulk degenerate electron gas and a small contamination of
accreting hot electrons. For this multicomponent astrophysical
system, we show that if, for a given equilibrium sequence, the total
energy is larger than some critical value (given in terms of invariant
helicities and the fractional coefficient of the hot component fraction),
the catastrophic loss of equilibrium could certainly occur.

For concrete boundary conditions, we will show analytically that the
catastrophe (brought about by slow changes in labels induced by
changing external conditions) pushes a Double Beltrami (DB) state
to relax to a minimum-energy single Beltrami field. During the
transition, much of the short--scale magnetic energy is converted
into the hot flow energy. For specific boundary conditions, the possibility
of the large--scale magnetic field formation is also explored explaining,
e.g,   the evolution of binaries, specifically the system of a dense/degenerate
WD’s outer layer that accretes classical hot astrophysical flow.

It is interesting to find that the initial state (energy, helicity
invariant values and  boundary conditions) contain much of the
information that holds the key to the eventual fate of a given
structure -- whether the structure maintains its integrity when
the surroundings undergo slow changes.

\bigskip

\section{Model Equations}

We study a quasi-neutral plasma consisting of a mobile
classical ion (i) component, and two relativistic electron
components -- the bulk degenerate ($d$) electron gas with a density
$N_{0d}$, and a small contamination of hot ($h$) electrons
with density $N_{0h}$. The quasi-neutrality condition can be
written as
\begin{equation}
N_{0d}+N_{0h}=N_{0i} \Rightarrow \frac{N_{0i}}{N_{0d}}=1
+ \alpha, \quad \alpha\equiv\frac{N_{0h}}{N_{0d}},
\label{DSS_1}
\end{equation}
where $\alpha\ll 1$ measures the extent of hot
electron contamination.

It was shown in \citep{SMB_2T} that the small
hot electron contamination, providing a new scale-length,
adds to the diversity in the scale-hierarchy of multi-component
plasmas met in astrophysical conditions. In present study, concentrating
on a special class of equilibria known as the Beltrami-Bernoulli (BB)
states, we explore the new channel for explosive/eruptive energy
transformations in such a mixture of relativistic plasmas often
emerging while the evolution of accreting stars / binaries.

By following Shatashvili et al (2019), one can deduce
(from the equations of motion) the  following dimensionless BB
equilibrium conditions for $d$ and $h$ electron components:
\begin{equation}
\boldsymbol B-\nabla\times(G_d\gamma_d\boldsymbol V_d)
= a_d\frac{n_d}{G_d}(G_d\gamma_d\boldsymbol V_d) \ ,
\label{DSS_12}
\end{equation}
\begin{equation}
\boldsymbol B-\nabla\times(G_h\gamma_h\boldsymbol V_h)
= \alpha a_h\frac{n_h}{G_h}(G_h\gamma_h\boldsymbol V_h).
\label{DSS_13}
\end{equation}
These Beltrami conditions align the generalized (canonical) vorticities
$\boldsymbol\Omega_{d(h)} = - \boldsymbol{B}
+\nabla\times(G_{d(h)}\gamma_{d(h)} \boldsymbol{V}_{d(h)})$
along their respective velocity fields. The overall force
balance demands that the Beltrami conditions must impose the
generalized Bernoulli Conditions (on electron fluids),
\begin{equation}
\nabla(G_d\gamma_d - \phi) = 0 \ , \qquad \qquad
\nabla(G_h\gamma_h - \phi) = 0 \ ,
\label{DSS_15}
\end{equation}
where $\phi$ is the electrostatic potential (of purely electromagnetic nature);
$n_{d(h)}=N_{d(h)}/\gamma_{d(h)}$
is the rest-frame particle density of the degenerate
(hot) electron fluid element ($N_{d(h)}$ being the laboratory frame density),
$V_{d(h)}$ is the fluid velocity, $\gamma_{d(h)}=(1-V^2_{d(h)}/c^2)^{-1/2}$.

The appearance of the constants $a_{d(h)}$ (Beltrami parameters)
is a reminder that the BB equilibria were derived from a variation principle
minimizing the system energy with helicity constraints; in fact these are
the Lagrange multipliers in the minimization process. The conserved
helicities (for each component) are defined by
\begin{equation}
h_{d(h)}=\int(\nabla^{-1}\times\boldsymbol\Omega_{d(h)})\cdot\boldsymbol
\Omega_{d(h)}d\boldsymbol r.
\label{DSS_18}
\end{equation}

The effective masses $G_d$ and $G_h$, occurring in the Beltrami conditions,
are quite different for the two electron species:
$G_d=\omega_d/n_d m_e c^2$ originates from degeneracy
equilibrium distribution function \\
\citep{Russo-2}
smoothly transfers to \\
$\omega_d=\omega_d(n)$ for a strongly degenerate electron plasma);
$\omega_d/n_dm_ec^2=(1+(R_d)^2)^{1/2}$, where $\omega_d$
is an enthalpy per unit volume; $R_d=(n_d/n_c)^{1/3}$
with $n_c=5.9\times 10^{29} cm^3$ being the critical number-density.
Then, the effective mass factor is determined by just the plasma
rest frame density, $G_d=[1+(n_d/n_c)^{2/3}]^{1/2}$ for an arbitrary $n_d/n_c$.
For relativistically hot plasma an expression for effective mass factor
$G_h$ can be found in \citep{BM-94,Ryu}.
These equations shall be coupled with ion fluid Beltrami Condition:
\begin{equation}
\boldsymbol{B}+\xi\nabla\times\boldsymbol{V}_i=(1+\alpha)a_i n_i\boldsymbol{V}_i \ ,
\quad \xi = [G^d_0\frac{m^d_e}{m_i}]^{-1} \ ,
\label{DSS_17}
\end{equation}
where $a_i$ is a Beltrami parameter related to ion--fluid
helicity $h_i$.

This set, together with Ampere's law
\begin{equation}
\nabla\times\boldsymbol B=\left[(1+\alpha)\boldsymbol{V}_i
-\boldsymbol V_d -\alpha\boldsymbol V_h   \right]
\label{DSS_19}
\end{equation}
defines the BB equilibrium states accessible to our astrophysical fluid
of two relativistic electron ($d$ and $h$) and one ion (i) components

Assuming quasi--neutrality to hold throughout the overall
incompressible dynamics, we put $\phi \equiv 0$; gravity and rotation
will be ignored for the time being like in \citep{SMB_2T}.

The following normalizations are used in the equations above: the density is
normalized to $N_{0d}$ (the corresponding rest-frame density is $n_{0d}$);
the magnetic field is normalized to some ambient measure $|\boldsymbol B_0|$;
hot electron gas temperature is normalized to $m_ec^2$; all velocities
are measured in terms of the corresponding Alfv\'en speed
$\boldsymbol V_A=\boldsymbol V_{Ad}=\boldsymbol B_0/\sqrt{4\pi n_{0d}m_e G_{0d}}$;
all lengths [times] are normalized to the ``effective'' degenerate electron skin
depth $\lambda_{eff}^d \ [\lambda_{eff}^d/\boldsymbol V_A]$, where
\begin{equation}
\lambda_{eff}^d = \frac{c}{\omega_{pe}^d}
= c\sqrt{\frac{m_eG_{0d}}{4\pi n_{0d} e^2}}
=\sqrt{\frac{\alpha G_{0d}}{G_{0h}}}\lambda_{eff}^h \ ,
\label{DSS_7}
\end{equation}
with $\lambda_{eff}^h=c\sqrt{\frac{m_e G_{0h}}{4\pi n_{0d}e^2}}$ \ ,
\begin{equation}
G_{0d}(n_{0d})=[1+R_{0d}^2]^{1/2}, \quad
R_{0d}=\bigg( \frac{n_{0d}}{n_c} \bigg)^{1/3} \ ,
\label{DSS_10}
\end{equation}
\begin{equation}
{\rm while} \qquad G_{0h}=\frac{5}{2}\frac{T_{e0}}{m_ec^2}+\frac{3}{2}
\sqrt{\left(\frac{T_{e0}}{m_ec^2}\right)+\frac{4}{9}} \ .
\label{DSS_11}
\end{equation}

Notice that there are two symmetry breaking mechanisms in
present model (each one being responsible for creating a net “current”):
1) the  $d$ and $h$ electrons have different effective inertias,
and 2) $h$ is a small contamination to the bulk $d$ electrons ($\alpha \ll 1$).
These are, in reality, different plasma species contributing two conserved
helicities that, eventually, translates into a higher index Beltrami state
(see \citep{lingam,SMB_multi} and references therein).

\section{Quadruple Beltrami Fields}

Let us first study a simple structure sustained by the equilibrium
equations displayed in the preceding section. In addition to
assuming $\phi\equiv 0$, let us put $\gamma_d\equiv 1$,
$\gamma_h\equiv 1$. The latter reduces the Bernoulli Conditions
({\ref{DSS_15}}) to $G_d = const= G_0 $; $G_h = const = H_0$.

In terms of the bulk-flow velocity
\begin{equation}
\mathbf{V} = \frac{1}{2}((1+\alpha)\mathbf{V}_i+\mathbf{V}_d),
\label{DSS_20}
\end{equation}
and $V_h$, we could write the ion and $d$ electron velocities as
\begin{equation}
\mathbf{V}_i = \frac{1}{1+\alpha}(\mathbf{V}+\frac{1}{2}\
\nabla\times\mathbf{B}+\frac{\alpha}{2}\mathbf{V}_h) \ ,
\label{DSS_21}
\end{equation}
\begin{equation}
\mathbf{V_d}=\mathbf{V}-\frac{1}{2}\nabla\times\mathbf{B}
- \frac{\alpha}{2}\mathbf{V}_h \ ,
\label{DSS_22}
\end{equation}
After straightforward algebra, we find that the equilibrium set of
equations can be reduced to single equation in $\mathbf{V}_h$
(see the details in \citep{SMB_2T}),
\begin{equation}
G_0 H_0 \nabla\times\nabla\times\nabla\times\nabla\times\mathbf{V}_h \ +
\label{DSS_30}
\end{equation}
\[
+ \ (\alpha a_h G_0 + a_1 H_0)\ \nabla\times\nabla\times\nabla\times\mathbf{V}_h \ +
\]
\[
+ \ (\alpha a_h a_1 + H_0 a_2 + \alpha H_0)\ \nabla\times\nabla\times\mathbf{V}_h \ +
\]
\[
+ \ (\alpha a_h a_2 - H_0a_3  - \alpha a_4)\ \nabla\times\mathbf{V}_h \ -
\]
\[
- \ \alpha(a_h a_3 + a_5)\mathbf{V}_h \ = \ 0 \ ,
\]
where
\[
a_1 = a_d -a_i(1+\alpha)\beta \ ,
\]
\[
a_2 = 1 + \beta (1+\alpha) - a_d a_i (1+\alpha) \beta G_0^{-1} \ ,
\]
\[
a_3 = \beta G_0{-1} (1+\alpha)(a_i-a_d) \ ,
\]
\[
a_4 = a_i(1+\alpha)\beta - a_d = G_0^{-1}[\eta^{-1}-a_d(1+\beta(1+\alpha))] \ ,
\]
\begin{equation}
a_5 = a_d a_i (1+\alpha)\beta G_0^{-1} \
\label{DSS_28}
\end{equation}
with
\begin{equation}
\notag
\eta=[a_i(1+\alpha)\beta+a_d]^{-1}, \quad \beta = \frac{G_0}{\xi} \ .
\end{equation}
what can be called a Quadruple Beltrami (QB) equation; the highest
derivative has four {\it curl} operators and all coefficients are
constants.

In terms of a set of obvious constants,
\begin{equation}
    \begin{aligned}
        b_1 &=G_0 \alpha a_h + a_1 H_0 \ , \\
        b_2 &=a_1 \alpha a_h + a_2 H_0 + \alpha G_0 \ , \\
        b_3 &= -a_2 \alpha a_h + a_3 H_0 + \alpha a_4 \ , \\
        b_4 &= a_3 a_h + a_5
    \end{aligned}
\label{DSS_31}
\end{equation}
equation (\ref{DSS_30}) takes the more compact form
\[
G_0 H_0 \nabla\times\nabla\times\nabla\times\nabla\times\mathbf{V}_h \
+ \ b_1 \nabla\times\nabla\times\nabla\times\mathbf{V}_h \ +
\]
\begin{equation}
+ \ b_2 \nabla\times\nabla\times\mathbf{V}_h \
- \ b_3 \nabla\times\mathbf{V}_h - \alpha b_4 \mathbf{V}_h \ = \ 0 \ ,
\label{DSS_32}
\end{equation}
that may be factorized as ($\nabla\times \equiv {\rm {\it curl}} $):
\begin{equation}
    (curl - \mu_1)(curl - \mu_2)(curl - \mu_3)(curl - \mu_4)\mathbf{V}_h = 0 \ .
\label{DSS_33}
\end{equation}


\begin{figure}
\begin{center}
\includegraphics[scale=0.5,angle=0]{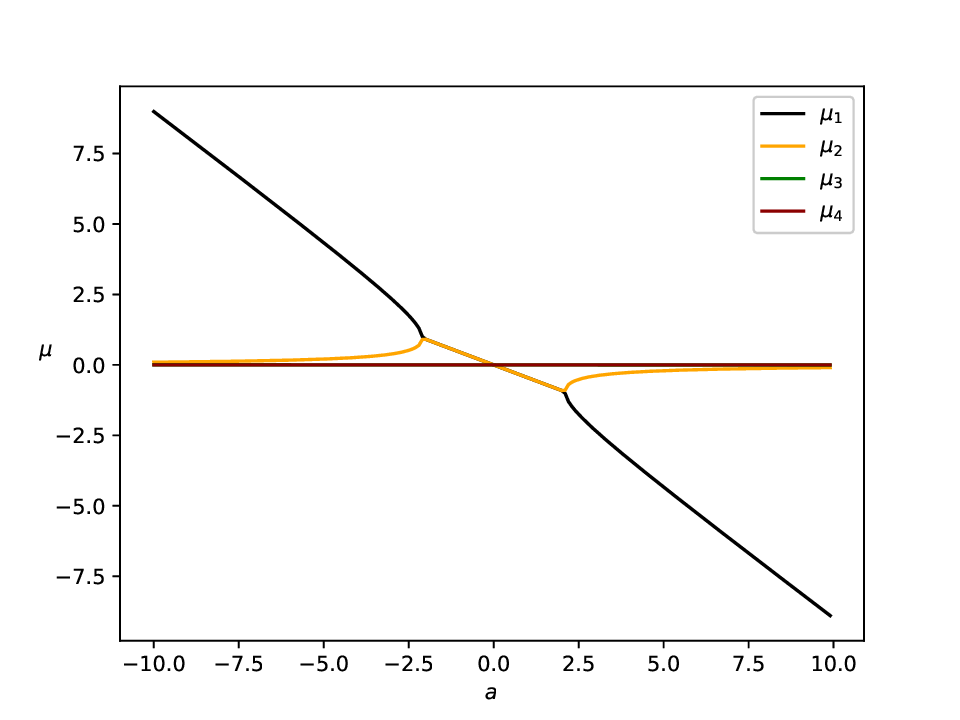}
\caption{Plot for the roots (inverse length-scales) of the equation (\ref{DSS_34}) for
$H_0 = 10 , \ \alpha = 10^{-3}$. The scale separation is clearly seen at Beltrami
parameter $a < -2.5$ and $a > 2.5$; non-zero roots being significantly
smaller ($\mu_3 \sim 10^{-3}$ and $\mu_4 \sim 10^{-20}$) are not well-distinguished.}
\label{Fig.1}
\end{center}
\end{figure}

The general solution of Eq. (\ref{DSS_32}) is a sum of four Beltrami
fields $\mathbf{F}_k$, each a solution of the
fundamental Beltrami Equation $\nabla\times\mathbf{F}_k =
\mu_k \mathbf{F}_k$. The  eigenvalues ( $\mu_k $) of the ${\it curl}$ operator
represent 4 inverse length scales associated with the system,
and are the solutions of the quartic equation
\begin{equation}
\mu^4+b_1^{*}\mu_3 +b_2^{*}-b_3^{*}\mu-b_4^{*} = 0 \ ,
\label{DSS_34}
\end{equation}
where
\begin{equation}
\begin{aligned}
b_1^{*} &=(G_0 H_0)^{-1} b_1 \ , \quad \quad b_2^{*} = (G_0 H_0)^{-1} b_2 \ ,\\
b_3^{*} &=(G_0 H_0)^{-1} b_3 \ , \quad \quad b_4^{*} = (G_0 H_0)^{-1} \alpha b_4.
\end{aligned}
\label{DSS_35}
\end{equation}
In Fig.1, we display the solutions of Eq. (\ref{DSS_34}) versus the Beltrami
parameter $a$ ($\equiv a_d \sim a_h$) when the hot electron fraction $\alpha = 10^{-3}$.
For this simple case, the following analytic formulas for the
$b$ coefficients in (\ref{DSS_28}) pertain:
\[
b_1 =G_0 \alpha a_h + (a_d - a_i \beta) H_0 \ ,
\]
\[
b_2 = (a_d - a_i \beta) \alpha a_h + (1+\beta-a_d a_i \beta G_0^{-1})H_0 + \alpha G_0 \ ,
\]
\[
b_3 =-\alpha a_h (1+\beta-a_d a_i \beta G_0^{-1}) + H_0\beta G_0^{-1}(a_i - a_d) +
\]
\[
\qquad \qquad     + \alpha(\beta a_i - a_d) \ ,\\
\]
\begin{equation}
b_4 = a_h \beta G_0^{-1}(a_i -a_d) + a_d a_i \beta G_0^{-1} \ .
\label{DSS_36}
\end{equation}

The general solution of (\ref{DSS_33}) will have the form
($C_{1,2,3,4}$ are arbitrary constants)
\begin{equation}
    \begin{aligned}
       & \mathbf{V}_h = C_1\mathbf{F}_1 + C_2\mathbf{F}_2 + C_3\mathbf{F}_3 + C_4\mathbf{F}_4 \ ,\\
       & \mathbf{B} = C_1^{'}\mathbf{F}_1 + C_2^{'}\mathbf{F}_2 + C_3^{'}\mathbf{F}_3 + C_4^{'}\mathbf{F}_4 \ ,\\
       & \mathbf{V}_d = C_1^{''}\mathbf{F}_1 + C_2^{''}\mathbf{F}_2 + C_3^{''}\mathbf{F}_3 + C_4^{''}\mathbf{F}_4 \ ,\\
       & \mathbf{V}_i = C_1^{'''}\mathbf{F}_1 + C_2^{'''}\mathbf{F}_2 + C_3^{'''}\mathbf{F}_3 + C_4^{'''}\mathbf{F}_4 \ ,
    \end{aligned}
\label{DSS_37}
\end{equation}
where
\[
C_{1,2,3,4}^{'} = (\alpha a_h + H_0 \mu_{1,2,3,4})C_{1,2,3,4} \ ,
\]
\[
C_{1,2,3,4}^{''} = \eta [G_0 (C_{1,2,3,4}^{'})\mu_{1,2,3,4}^{2}
\]
\[
\qquad - \ a_i(1+\alpha)\beta(C_{1,2,3,4}^{'})\mu_{1,2,3,4} \ +
\]
\begin{equation}
\qquad + \ (1+\beta(1+ \alpha))(C_{1,2,3,4}^{'}) \ +
\label{DSS_38}
\end{equation}
\[
\qquad + \ \alpha G_0 \mu_{1,2,3,4} \ - \ \alpha a_i (1+\alpha)\beta]C_{1,2,3,4} \ ,
\]
$$
C_{1,2,3,4}^{'''} = \frac{\eta}{1+\alpha}[G_0(C_{1,2,3,4}^{'})\mu_{1,2,3,4}^2
+ a_d(C_{1,2,3,4}^{'})\mu_{1,2,3,4}
$$
$$
+(1+\beta(1+\alpha)) (C_{1,2,3,4}^{'}) + \alpha G_0\mu_{1,2,3,4} + \alpha a_d]C_{1,2,3,4} \ .
$$
Hence, a small contamination of hot electrons made the
structure--hierarchy (QB states) richer as compared to Double--Beltrami
equilibrium states. Let us now examine if explosive--eruptive
phenomena are possible in such a composite system.

\section{Analysis for catastrophic transformations}

As mentioned in the introduction, the main goal of present work is to
explore conditions for the catastrophic energy transformations  accompanied
by the generation of macro--scale fields. For analytic simplicity,
we consider a specific case for our composite three-fluid system:
in order to highlight the effects of (the small contamination) of hot electron
fluid, we choose our coefficients to make the last two terms in (\ref{DSS_32})
go to zero reducing the dimensionality (measuring the number of {\it curl} operators)
by two. Thus, the Quadruple Beltrami system reduces to a Double Beltrami (DB)
system (\ref{DSS_30}). We notice that
\[
\qquad b_4 = 0, \quad \Rightarrow \qquad
a_i (a_d  + a_h ) = a_h a_d \ ,
\]
from where we get the additional condition for Beltrami parameters
(see \citep{SMB_2T} for emerging scale hierarchy):
\begin{equation}
\qquad a_i = \frac{a}{2} \qquad {\rm{if}} \qquad a_d \sim a_h \equiv a \ ,
\label{DSS_39}
\end{equation}
application of which reduces Eq.(\ref{DSS_30}) to the triple
Beltrami (TB) equation. In addition,
\[
b_3 = 0 \quad \Rightarrow
\quad H_0\beta G_0^{-1}(a_i - a_d)\ + \ \alpha(\beta a_i - a_d) \ -
\]
\[
\qquad \qquad \qquad - \ \alpha a_h(\beta +1 \ - \ a_d a_i \beta G_0^{-1}) \ = \ 0
\]
which, due to (\ref{DSS_39}), yields:
\[
\qquad \qquad \frac{H_0}{\alpha} = a^2 - G_0\left(1+\frac{4}{\beta}\right)
\]
leading to the final condition for $a^2$ linking the Beltrami parameters to the
physical parameters defining the system ( $\alpha \ll 1\sim 10^{-3}$) :
\begin{equation}
a^2 = \frac{H_0}{\alpha} + G_0\left(1+\frac{4}{\beta}\right) \ .
\label{DSS_40}
\end{equation}

For a realistic choice for effective masses  \\
\cite{KS_2TRD} -- $H_0 \geq 10$, and $\quad 2.5 > G_0 \geq 1.1$  -- we obtain:
\[
\qquad \frac{H_0}{\alpha} \gg G_0\left(1+\frac{4}{\beta}\right)
\label{DSS_41}
\]
leading to
\begin{equation}
\qquad a^2 \simeq \frac{H_0}{\alpha} \
\label{DSS_42}
\end{equation}
With the use of (\ref{DSS_39}) and (\ref{DSS_42}) conditions,
Eq.(\ref{DSS_32}) is reduced to the DB equation.

In order to extract the special role of the hot electron contamination,
it is useful to assume $G_0 \gtrsim 1$ (ignoring the effects of degeneracy
for $d$ electrons) which allows us to neglect the the inertial term
in Eq.(\ref{DSS_12}).  The Beltrami conditions
for all 3 fluids, then, simplify to:
\begin{equation}
\mathbf{B} = a \mathbf{V}_d \ ,
\label{DSS_43}
\end{equation}
\begin{equation}
\mathbf{B} - H_0 \nabla\times\mathbf{V}_h = \alpha a \mathbf{V}_h \ ,
 \label{DSS_44}
\end{equation}
\begin{equation}
 \mathbf{B}+\xi \nabla\times\mathbf{V}_i = (1+\alpha)\frac{a}{2}\mathbf{V}_i \ .
 \label{DSS_45}
\end{equation}
that will lead to the same DB equation.

Using Eq.(\ref{DSS_43}), the bulk ``flow velocity'' (Eq.(\ref{DSS_20}))
is expressible as:
\begin{equation}
\mathbf{V} = \frac{1}{2} \nabla\times\mathbf{B}+\frac{1}{a}\mathbf{B}
+ \frac{\alpha}{2}\mathbf{V}_h \ .
\label{DSS_46}
\end{equation}
The preceding equations, along with Ampere's law (\ref{DSS_19}),
yield, after some straightforward algebra,  expressions for the velocity--fields of
$h$ electron flow and ion-=flow
\begin{equation}
\mathbf{V}_h = \frac{2}{(\alpha + \frac{H_0}{\xi})}
\left(\left(\frac{a}{2 \xi} - \frac{1}{a}\right)\mathbf{B} -
 \nabla\times\mathbf{B}\right) \ ,
\label{DSS_50}
\end{equation}
\[
\mathbf{V}_i = \frac{1}{1+\alpha} \nabla\times\mathbf{B}
\ + \ \frac{1}{2}\ \mathbf{B} \ +
\]
\begin{equation}
\qquad \quad + \ \frac{2\alpha}{(\alpha +\frac{H_0}{\xi})}
\left(\left(\frac{a}{2 \xi} - \frac{1}{a}\right)\mathbf{B}
- \nabla\times\mathbf{B}\right) \ ,
\label{DSS_51}
\end{equation}
and finally, the DB equation:
\[
\nabla\times\nabla\times\mathbf{B} \ - \ \left(\frac{a}{2 \xi}
- \frac{1}{a} - a \frac{\alpha}{H_0}\right)\nabla\times\mathbf{B} \ +
\]
\begin{equation}
+ \ \frac{1}{2}\left(\frac{1}{\xi} - \frac{\alpha}{H_0}
\left(\frac{a^2}{\xi}-3\right)\right)\mathbf{B} = 0 \ .
\label{DSS_52}
\end{equation}
for the magnetic field.

Notice, that in the $\alpha \to 0$ limit,  the DB Eq.(\ref{DSS_52})
is structurally the same as for the pure electron--ion
plasma in \cite{osym,osym2}, and becomes exactly the same when $\alpha =0$.

One may now factorize Eq.(\ref{DSS_52})
\[
\qquad \qquad (curl-\mu_1)(curl - \mu_2)\mathbf{B}=0 \ ,
\]
where $\mu_{1,2}$, determined as
 \begin{equation}
\mu^2 - b^{'}_1 \mu +b^{'}_2\  = \ 0 \ ,
\label{DSS_53}
\end{equation}
with ($\Tilde{a} = a^{-1}$),
\[
b^{'}_1 = \frac{a}{2 \xi} - \Tilde{a}-\frac{\alpha}{H_0} \ ,
\quad b^{'}_2 = \frac{1}{2}\left(\frac{1}{\xi}
- \frac{\alpha}{H_0}\left(\frac{a^2}{\xi}-3\right)\right)  ,
\]
are the eigenvalues of the $curl$  operator and represent the two
inverse length scales of the system. The eigenvalues $\mu_{1,2}$ have
explicit expressions in terms of system parameters
\begin{equation}
\qquad \qquad \mu_{1,2} \ = \ \frac{1}{2}\left(\frac{a}{\xi} -
\Tilde{a}-a\frac{\alpha}{H_0}\right) \ \pm
\label{DSS_54}
\end{equation}
\[
\pm \frac{1}{2} \sqrt{\left(\frac{a}{\xi}-\Tilde{a}
-a \frac{\alpha}{H_0}\right)^2-2\left(\frac{1}{\xi} -
\frac{\alpha}{H_0}\left(\frac{a^2}{\xi}-3\right)\right)} \ ,
\]
and obey the following identities
\begin{equation}
\mu_1 + \mu_2 = \frac{a}{2 \xi} - \Tilde{a} - a \frac{\alpha}{H_0} \ ,
\label{DSS_55}
\end{equation}
\begin{equation}
\mu_1 \mu_2 = \frac{1}{2}\left(\frac{1}{\xi}-\frac{\alpha}{H_0}
\left(\frac{a^2}{\xi}-3\right)\right) \ ,
\label{DSS_56}
\end{equation}
\begin{equation}
(\mu_1 + \Tilde{a})(\mu_2+\Tilde{a}) = \frac{1}{\xi}
- \frac{\alpha}{2 H_0}\left(\frac{a^2}{\xi} - 1\right) \ .
\label{DSS_57}
\end{equation}


\begin{figure}
\begin{center}
\includegraphics[scale=0.5,angle=0]{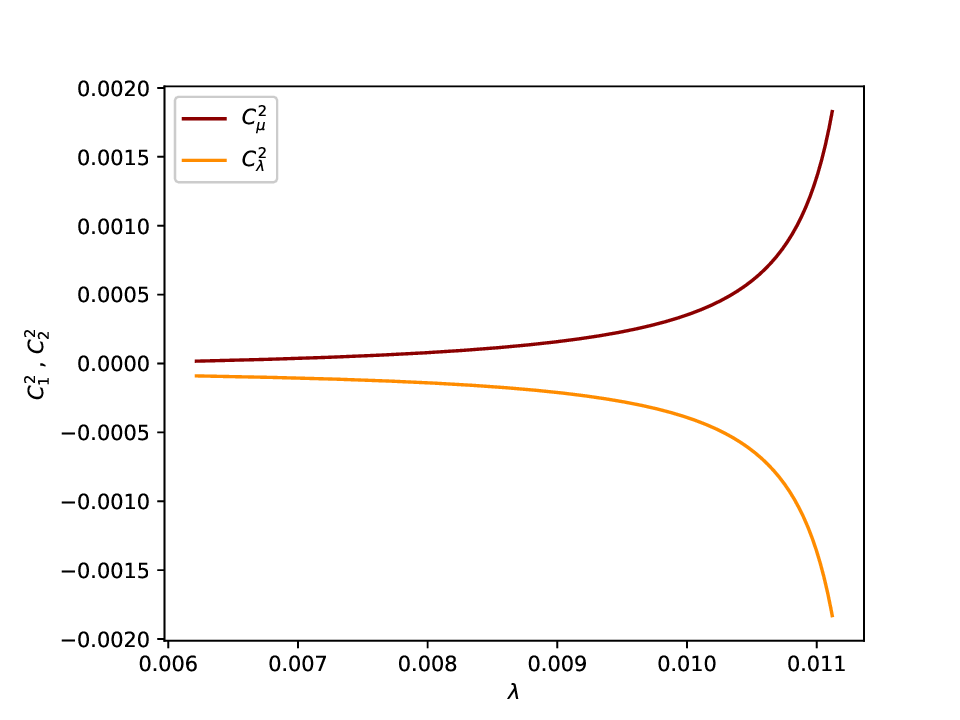}
\caption{Plots for the amplitudes $C_1$ and $C_2$ versus $\lambda $
when applying the conditions (\ref{DSS_39}-\ref{DSS_42}) leading
to the Double Beltrami equation (\ref{DSS_52}) for the
specific physical parameters: $a = [-48, -35.2]; \ h_d = - 25; \
h_h = -40; \ E = 5;  \ E_{crit} = 2.1$, \ $H_0 = 10 , \ \alpha = 10^{-3}$. }
\label{Fig.2.}
\end{center}
\end{figure}


\begin{figure}
\begin{center}
\includegraphics[scale=0.5,angle=0]{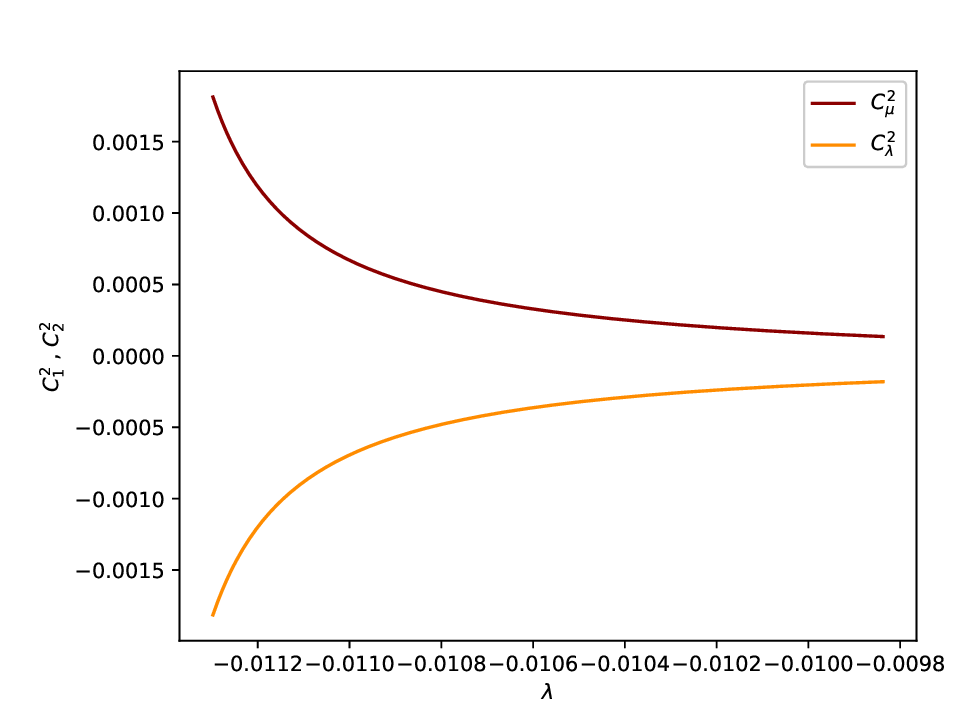}
\caption{Plots for the amplitudes $C_1$ and $C_2$ versus $\lambda $
when applying the conditions (\ref{DSS_39}-\ref{DSS_42}) leading
to the Double Beltrami equation (\ref{DSS_52}) for the
specific parameters: \ $a = [34.8, 38]; \ h_d = 40;  \ h_h = 25; \
E = 5; \ E_{crit} = 2.1$, \ $H_0 = 10 , \ \alpha = 10^{-3}$.}
\label{Fig.3.}
\end{center}
\end{figure}

If $\mathbf{G_1}$ and $\mathbf{G_2}$ are the Beltrami
eigenvectors associated with the eigenvalues $\mu_{1}, \mu_{2}$,
the general DB solution ( for all relevant physical variables) may be
constructed as ($C_{1,2}$ are arbitrary constants (amplitudes)):
\begin{equation}
\mathbf{B} = C_1 \mathbf{G_1} + C_2 \mathbf{G_2} \ ,
\label{DSS_58}
\end{equation}
\begin{equation}
\mathbf{V_d} = \Tilde{a} C_1 \mathbf{G_1} + \Tilde{a} C_2 \mathbf{G_2} \ ,
\label{DSS_59}
\end{equation}
\begin{equation}
\mathbf{V_h} = C_1^{'} \mathbf{G_1} + C_2^{'} \mathbf{G_2} \ ,
\label{DSS_60}
\end{equation}
\begin{equation}
\mathbf{V_i} = C_1^{''} \mathbf{G_1} + C_2^{''}\mathbf{G_2} \ ,
\label{DSS_61}
\end{equation}
with the constants
\begin{equation}
C_{1,2}^{'} = \frac{1}{\alpha(1+\frac{H_0}{\alpha \xi})}
\left(\left(\frac{a}{\xi}-2\Tilde{a}\right) -
2\mu_{1,2}\right) C_{1,2} \ ,
\label{DSS_62}
\end{equation}
\begin{equation}
C_{1,2}^{''} = \left(\mu_{1,2} + \Tilde{a} +
\frac{1}{1+\frac{H_0}{\alpha \xi}}\left(\frac{a}{\xi}-2\Tilde{a
} - 2\mu_{1,2}\right)\right) {C}_{1,2} \ .
\label{DSS_63}
\end{equation}


\begin{figure}
\begin{center}
\includegraphics[scale=0.5,angle=0]{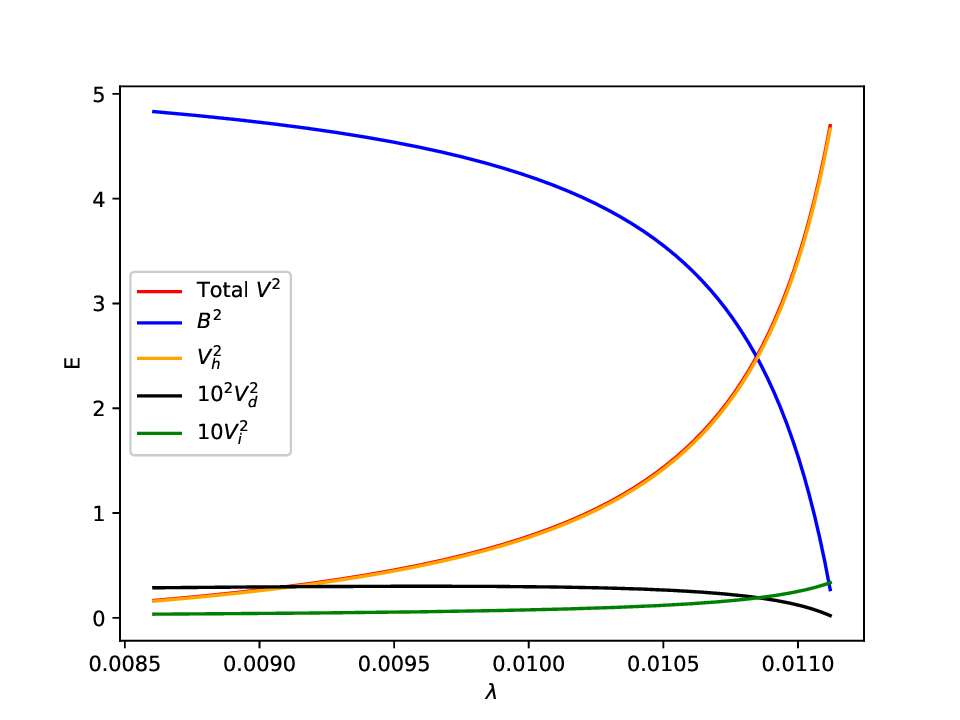}
\caption{Plots for the magnetic and fluid energies versus $\lambda $
for different species for the case 1 (presented in Figure 2):
total fluid energy (red) is dominated
by the hot fraction fluid energy. For given parameters the magnetic
field (blue) energy is converted to flow energy at the catastrophe. }
\label{Fig.4.}
\end{center}
\end{figure}


\begin{figure}
\begin{center}
\includegraphics[scale=0.5,angle=0]{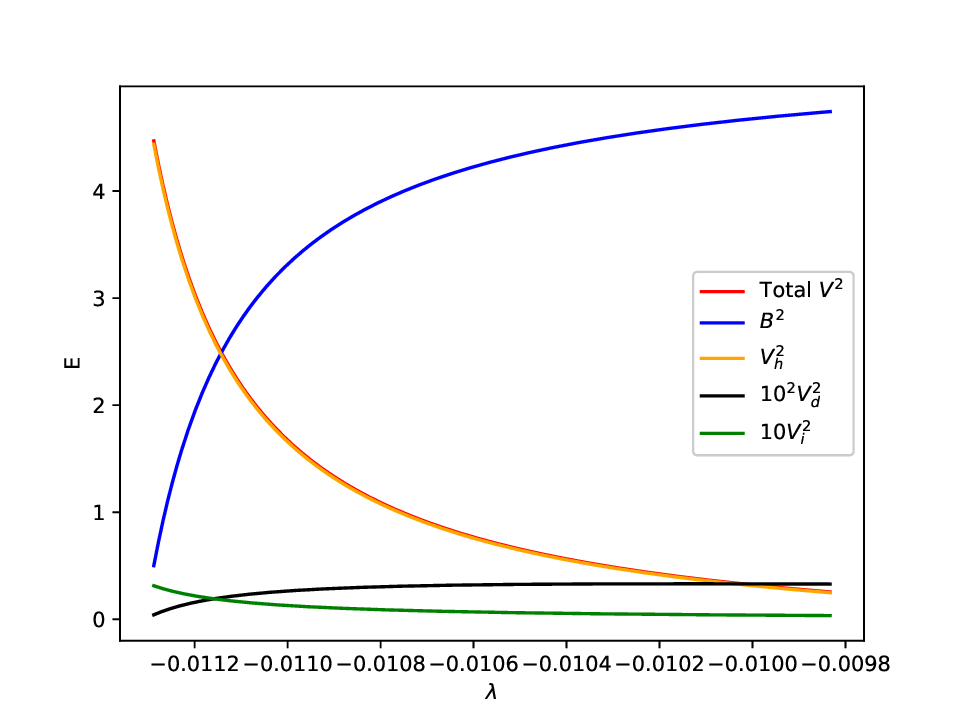}
\caption{Plots for the magnetic and fluid energies versus $\lambda $
for different species for the case 2 (presented in Figure 3):
total fluid energy (red) is dominated by
the hot fraction fluid energy. For given parameters the total fluid energy
is converted to magnetic field energy (blue) at the catastrophe.}
\label{Fig.5.}
\end{center}
\end{figure}

\subsection{Conservation laws}

In the general vortex dynamics, the helicities (\ref{DSS_18}) of
each fluid are conserved. The three conserved helicities correspond
to the generalized vorticities
\begin{equation}
\mathbf{\Omega}_{d(h)} = -\mathbf{B} + \nabla\times(G_{d(h)}\mathbf{V}_{d(h)}) ,
\label{DSS_65}
\end{equation}
\begin{equation}
\mathbf{\Omega}_i = \mathbf{B} + \xi \nabla \times \mathbf{V}_i.
\label{DSS_66}
\end{equation}

An additional constant of motion is the total energy,
\begin{equation}
E=\frac{1}{2}\int(\mathbf{B}^2 + \alpha \mathbf{V}_h^2 +\mathbf{V}_d^2
+ (1+\alpha)\mathbf{V}_i^2)  \ d{\bf r}.
\label{DSS_67}
\end{equation}

From the vorticities ($\mathbf{\Omega}_{d,h,i}$ and
the BB solutions we may, explicitly, compute the helicities and energy}
($L$ is the size of the system):

\noindent 1) degenerate electron helicity,
\begin{equation}
h_d = \frac{L^2}{2}\left(\frac{C_1^2}{\mu_1}+\frac{C_2^2}{\mu_2}\right) \ ,
\label{DSS_69}
\end{equation}
\noindent 2) hot electron helicity,
\[
h_h = \frac{L^2}{2}\frac{1}{\mu_1}\left(1 - l_1\mu_1\frac{H_0}{\alpha}\right)^2C_1^2
\]
\begin{equation}
\qquad + \ \frac{L^2}{2}\frac{1}{\mu_2}\left(1 - l_2\mu_2\frac{H_0}{\alpha}\right)^2C_2^2
 \label{DSS_70}
\end{equation}
\noindent 3) the ion fluid helicity,
\[
h_i = \frac{L^2}{2}\frac{1}{\mu_1}(1+\xi\mu_1(\mu_1 + \Tilde{a}+l_1)^2)C_1^2
\]
\begin{equation}
\qquad + \ \frac{L^2}{2}\frac{1}{\mu_2}(1+\xi\mu_2(\mu_2 + \Tilde{a}+l_2)^2)C_2^2 \ ,
\label{DSS_71}
\end{equation}
and

\noindent 4)the total energy:
\[
E = \frac{L^2}{2}\left(1 + \frac{l_1^2}{\alpha}+\Tilde{a}^2+(\mu_1 
+ \Tilde{a} + l_1)^2\right)C_1^2
\]
\begin{equation}
\qquad + \ \frac{L^2}{2}\left(1+\frac{l_2^2}{\alpha}+\Tilde{a}^2
+ (\mu_2+\Tilde{a}+l_2)^2\right)C_2^2 \ ,
\label{DSS_72}
\end{equation}
where
\[
l_1 = \frac{1}{1+\frac{H_0}{\alpha \xi}}\left(\frac{a}{\xi}-2\Tilde{a}-2 \mu_1\right) \ ,
\]
\begin{equation}
l_2 = \frac{1}{1+\frac{H_0}{\alpha \xi}}\left(\frac{a}{\xi}-2\Tilde{a}-2 \mu_2\right) \ ,
\label{DSS_68}
\end{equation}

Using (\ref{DSS_69})-(\ref{DSS_71}), we derive the expressions for
$\Tilde{h_1} = h_i - h_d$ and $\Tilde{h_2} = h_h-h_d$ , respectively:
\[
\Tilde{h_1} = h_i-h_d =
\]
\[
\qquad = \frac{L^2}{2}(2\xi(\mu_1+\Tilde{a}+l_1)+\xi^2\mu_1(\mu_1+\Tilde{a}+l_1)^2)C_1^2
\]
\begin{equation}
\qquad + \ \frac{L^2}{2}(2\xi(\mu_2+\Tilde{a}+l_2)+\xi^2\mu_2(\mu_2+\Tilde{a}+l_2)^2)C_2^2 \ ,
\label{DSS_73}
\end{equation}

\[
\Tilde{h_2} = h_h-h_d = \frac{L^2}{2}\,\frac{H_0}{\alpha}\,l_1\,\left(l_1^2 \mu_1
\frac{H_0}{\alpha} - 2\right)C_1^2
\]
\begin{equation}
\qquad + \ \frac{L^2}{2}\,\frac{H_0}{\alpha}\,l_2\,\left(l_2 \mu_2 \frac{H_0}{\alpha}
- 2\right)C_2^2 \ .
\label{DSS_74}
\end{equation}
As expected, when $\alpha \rightarrow 0$ (no contamination --
just two species) Eqs (\ref{DSS_72})-(\ref{DSS_73}) exactly coincide
with their counterparts derived in \citep{osym,osym2} for the
classical electron-ion fluid.

\subsection{Catastrophe Condition}

From the equations (\ref{DSS_69}) and (\ref{DSS_72}), we get
\begin{equation}
C_1^2 = \frac{1}{\tilde{D}}\frac{2\mu_1}{L^2}E\left(1-l_2\mu_2\frac{H_0}{\alpha}\right)^2 -
\label{DSS_75}
\end{equation}
\[
\qquad - \ \frac{1}{\tilde{D}}\frac{2\mu_1}{L^2}\,\mu_2 \,h_h
\left(1+\frac{l_2^2}{\alpha}+\tilde{a}^2+(\mu_2+\tilde{a}+l_2)^2\right) \ ,
\]
\begin{equation}
C_2^2 = - \frac{1}{\tilde{\tilde{D}}}\frac{2\mu_2}{L^2}E
\left((1-l_1\mu_1\frac{H_0}{\alpha}\right)^2  +
\label{DSS_76}
\end{equation}
\[
\qquad + \ \frac{1}{\tilde{D}}\frac{2\mu_2}{L^2} \mu_1 h_h \left(1+\frac{l_1^2}
{\alpha}+\tilde{a}^2+(\mu_1+\tilde{a}+l_1)^2\right) ,
\]
where
\[
D = \mu_1\left(1+\frac{l_1}{\alpha}+\Tilde{a}^2+(\mu_1+\Tilde{a}+l_1)^2\right) -
\]
\[
\qquad - \ \mu_2\left(1+\frac{l_2}{\alpha}+\Tilde{a}^2+(\mu_2+\Tilde{a}+l_2)^2\right) \ ,
\]
\[
\tilde{D} = \mu_1(1+\frac{l_1^2}{\alpha}+\tilde{a}^2
+ (\mu_1+\tilde{a}+l_1)^2)(1-l_2\mu_2\frac{H_0}{\alpha})^2 -
\]
\[
\qquad - \ \mu_2(1+\frac{l_2^2}{\alpha}+\tilde{a}^2
+ (\mu_2+\tilde{a}+l_2)^2)(1-l_1\mu_1\frac{H_0}{\alpha})^2 \ .
\]

Applying (\ref{DSS_42}), with $\alpha \ll 1$ one can rewrite
the relations (\ref{DSS_55})-(\ref{DSS_57}) as follows:
\begin{equation}
\mu_1+\mu_2 = \frac{a}{2 \xi} - 2\Tilde{a} \ , \qquad \mu_1 \mu_2
= \frac{3}{2} \frac{\alpha}{H_0} \ ,
\label{DSS_77}
\end{equation}
\begin{equation}
(\mu_1 + \Tilde{a})(\mu_2+\Tilde{a}) = \frac{1}{2 \xi}+\frac{\alpha}{2H_0}
\simeq \frac{1}{2 \xi} \ ,
\label{DSS_78}
\end{equation}
Also, for following effective masses: $H_0 \simeq 10$, $G_0 \simeq 1.1$
and for $\xi \simeq 2000$ leading to
\[
\qquad \qquad \qquad \frac{H_0}{\alpha \xi} \gg 1
\]
one can simplify the expressions for $l_1, l_2$ yeilding:
\begin{equation}
l_1 \simeq \frac{\alpha \xi}{H_0}(\frac{a}{\xi}-2 \Tilde{a}
- 2 \mu_1), \quad  l_2 \simeq \frac{\alpha \xi}{H_0}(\frac{a}{\xi}
- 2 \Tilde{a} - 2 \mu_2) \
\label{DSS_81}
\end{equation}
and, after simple straightforward algebra, we rewrite the equations
(\ref{DSS_72})-(\ref{DSS_73}) as follows:
\begin{equation}
\qquad \qquad E =
\label{DSS_82}
\end{equation}
\[ \frac{L^2}{2}\left(1+(\mu_1+2 \Tilde{a})^2+\Tilde{a}^2
+ \frac{\alpha\xi^2}{H_0^2}(\frac{1}{\xi}-2\Tilde{a}-2\mu_1)^2\right)C_1^2
\]
\[
+ \frac{L^2}{2}\left(1+(\mu_2+2 \Tilde{a})^2+\Tilde{a}^2
+ \frac{\alpha\xi^2}{H_0^2}(\frac{1}{\xi}-2\Tilde{a}-2\mu_2)^2\right)C_2^2  ,
\]
\[
\Tilde{h}_1 = \frac{L^2}{2}b[(1+\xi(\mu_1+2\Tilde{a})^2)C_1^2
+(1+\xi(\mu_2+2\Tilde{a})^2)C_2^2]
\]
\begin{equation}
\qquad - \ \xi\mu_1\mu_2h_d \ ,
\label{DSS_85}
\end{equation}
where, using Eq.(\ref{DSS_77}), we have for $a$ and $b$ following:
$$
a = 2\xi(\mu_1+\mu_2+2\Tilde{a}), \quad b=\frac{a}{2}
=\xi(\mu_1+\mu_2+2\Tilde{a}) \ .
$$

Using Eq. (\ref{DSS_82}) in Eq.(\ref{DSS_85}), we find
a link between energy and helicities:
\[
\Tilde{h}_1 =b\xi E -\xi\mu_1\mu_2 h_d + \frac{L^2}{2}
\big(b\xi(\frac{1}{\xi}-1)(C_1^2+C_2^2)\big)
\]
\[
\qquad - \ \frac{L^2}{2}\big(\alpha\frac{\xi^2}{H_0^2}(\frac{a}{\xi}
-2\Tilde{a}-2\mu_1)^2+\Tilde{a}^2\big)C_1^2
\]
\begin{equation}
\qquad - \ \frac{L^2}{2} \big(\alpha\frac{\xi^2}{H_0^2}(\frac{a}{\xi}
-2\Tilde{a}-2\mu_2)^2+\Tilde{a}^2\big)C_2^2 \ .
\label{DSS_86}
\end{equation}
Notice, that with no hot fraction ($\alpha \rightarrow 0$,  $\lambda_d
\rightarrow \sqrt{\xi}\lambda_i$), Eq.(\ref{DSS_86}) corresponds
to and exactly equals the expression derived for e-i fluid by
\cite{osym,osym2}. Then, Eq.(\ref{DSS_86}) can be rewritten as:
\[
\qquad \qquad \Tilde{h}_1 \sim bE-\mu_1\mu_2 h_d+\alpha f(h_h,\Tilde{a}) \ ,
\]
where $f$ is some function of hot electron helicity, defined
by $H_0$ and $\alpha$ parameters; here $\Tilde{a} = \sqrt{\frac{\alpha}{H_0}}$.
Using (\ref{DSS_81}) and (\ref{DSS_42})
we simplify the expression for $\Tilde{h}_2$:
\[
\Tilde{h}_2 = \frac{L^2}{2}\big(\ \mu_1 \xi^2(\frac{a}{\xi}-2\Tilde{a}-2\mu_1)^2
- 2\xi(\frac{a}{\xi}-2\Tilde{a}-2\mu_1) \big)C_1^2
\]
\begin{equation}
+ \frac{L^2}{2}\big(\ \mu_2 \xi^2(\frac{a}{\xi}-2\Tilde{a}-2\mu_2)^2
 - 2\xi(\frac{a}{\xi}-2\Tilde{a}-2\mu_2) \big)C_2^2 \ .
\label{DSS_87}
\end{equation}

After long and tedious algebra we simplify the expressions
for $\Tilde{h}_1$ and $\Tilde{h}_2$ and express them by the defining
system parameters. Below in the analysis $\lambda$ corresponds to
the macro--scale and $\mu$ to the micro--scale. If the curve
$\lambda(\mu)$ has an extremum, i.e., $d\lambda/d\mu=0$
for real $\lambda$ and $\mu$, then it implies the
disappearance of the micro-scale constituent of the DB field
and we can derive the conditions for the possibility
of a catastrophic rearrangement of the original state
(see details in \citep{osym2} for the case of
catastrophic transformation of DB state).
Since $\lambda$ and $\mu$ are fully determined in terms of $b$
and $\tilde{a}$, the extremum condition $d\lambda/d\mu=0$
may be replaced by $d\lambda/d\tilde{a}=0$. Then, using the
equation (\ref{DSS_54}), we find:
\[
\frac{d\lambda}{d\tilde{a}} = -1 + \frac{2\,(b\,{\xi}^{-1}
- \tilde{a})}{\sqrt{(\frac{2b}{\xi}-2\tilde{a})^2+6}}
\]
\begin{equation}
\qquad + \, \frac{1}{\xi}\frac{d b}{d \tilde{a}}\left(1-\frac{2\,(b\,{\xi}^{-1}
- \tilde{a})}{\sqrt{(\frac{2b}{\xi}-2\tilde{a})^2 + 6}}\right) = 0 \ ,
\label{DSS_96}
\end{equation}
from where we get:
\begin{equation}
\qquad \qquad \frac{db}{d\tilde{a}} = \xi \ .
\label{DSS_97}
\end{equation}
Using $\tilde{a}\ll1$, $\alpha\ll1$, $\xi \simeq 2000$,
we find:
\begin{equation}
b = - 4\frac{(h_i-h_d)\tilde{a}^3+2}{E\tilde{a}^2+\frac{1}{\alpha}
(h_h-h_d)\tilde{a}^3-\frac{1}{\alpha}4\xi(h_h-h_d)\tilde{a}^4} \
\label{DSS_104}
\end{equation}
and, then, we obtain for \ $db/d \tilde{a}$ \ following (using the
more simplified expression for $\Tilde{h}_2$):
\[
\frac{db}{d\tilde{a}}=-\frac{12 (h_i-h_d)}{E
+\frac{1}{\alpha}\tilde{a}(h_h-h_d)-\frac{1}{\alpha}4\tilde{a}^2\xi(h_h-h_d)} +
\]
\[
\qquad + \left(2+4\tilde{a}^3(h_i-h_d) \right)\cdot
\]
\begin{equation}
\qquad \cdot \,\frac{2 \tilde{a}E + \frac{3}{\alpha}\tilde{a}^2(h_h-h_d)
- \frac{16}{\alpha}{\tilde{a}}^3\xi(h_h-h_d)}
{\left(E\tilde{a}^2 + \frac{1}{\alpha} \tilde{a}^3(h_h-h_d)
-\frac{4}{\alpha}\tilde{a}^4\xi(h_h-h_d)\right)^2} \ ,
\label{DSS_105}
\end{equation}
which, using Eq.(\ref{DSS_97}), reduces to
(to the lowest order):
\begin{equation}
4E  + \frac{12}{\alpha}(h_h-h_d)\tilde{a}
-\frac{32}{\alpha}(h_h-h_d)\xi \tilde{a}^2 = 0 \
\label{DSS_107}
\end{equation}
Then $\tilde{a}$ is real (at $d\lambda/d\tilde{a}=0$) if
\begin{equation}
\frac{9}{\alpha^2}(h_h-h_d)^2+\frac{32}{\alpha}(h_h-h_d)\xi E \geq 0 \
\label{DSS_108}
\end{equation}
and from Eq. (\ref{DSS_108}) we conclude that the extremum is physical when:
\begin{equation}
E \geq E_{crit} = \frac{9}{32}\frac{h_d-h_h}{\alpha \xi} \ .
\end{equation}

We remind the reader, that the extremum condition ($d\lambda/d\mu=0$)
does represent a critical transition point; if the system is pushed
beyond this point, this will result in a loss of equilibrium.
Plots for amplitudes $C_1^2$ and $C_2^2$ versus $\lambda $
are presented in Figures 2,3 for the parameters of the
DB state under consideration.

Thus, we were able to show that our rather simple representative
system has a potential to undergo catastrophic transformation of energies;
in fact, we were successful in deriving conditions for which  such a
change could occur in the binary relativistic multi--component fluid.
We investigated in detail a simple (extreme parameter) system for which
the general Quadruple Beltrami system reduces to  the much simpler Double
Beltrami one. The conditions for a ``catastrophe'' help define scenarios
for generating  macro--scale velocity and/or magnetic fields. Plots in
Figures 4-5 show, that the total flow energy (being strongly Super-Alfv\'enic)
is carried mostly by the $h$ electrons. The clear message is that for
appropriate choice of initial conditions, it is possible to generate
strong macro--scale magnetic field of Fig.5 (velocity field of Fig.4).
In the former (latter) all of the flow  energy (magnetic energy) is
converted to the magnetic energy (kinetic energy) at the catastrophe.

We make here a diversion to point out that similar possibilities
were explored in \\
\cite{KS_2TRD} via applying the dynamic Unified Dynamo approach.
Studying  a realistic binary system of a WD accreting a hot astrophysical
flow, the formation of dispersive strong super--Alfv\'enic macro--scale
flow/outflow (Alfv\'en–-Mach number $> 10^6$),  and/or the generation
of super--strong magnetic fields was demonstrated. This approach, in fact,
is complementary to the successive equilibrium approach invoked in present paper.

Notice that the scenario developed in this paper, absent in a pure
magnetized degenerate e–-i plasma, emerges entirely due to the
hot contamination ($h$ electrons) observed in accreting stars/binary systems.

That creation of super--Alfv\'enic large--scale hot flows (found to be fed by
short--scale fluctuations of both fluids) in this composite system, could be
the mechanism behind  the formation of transient jets.

We emphasize here that, in present work, we have found another,
explosive path for the exploration of high-field Magnetic WDs in
binary systems (in Fig.5, we show that the total flow energy dominated
by the hot fraction flow energy is converted to the magnetic-field energy
at the catastrophe) in conformity with the recent argument:
1) that the formation of such WDs are related to the binary
interactions during the post-main-sequence phases of star
evolutions (see e.g., Nordhaus et al (2010)),
2) that many stars are born in the binary systems
going through one or more phases of the mass exchange
\citep{winget,DAZ,tremblay}; \\
\citep{mukai}. Such scenarios could also pertain to
magnetically induced stellar outbursts in WD binaries
\cite{Qian}.

In addition we explored the explosive path for the formation
of strong large--scale flows (see Fig.4) when the magnetic energy
is fully converted to the flow energy at the catastrophe.
Such transient flows could be additional sources for
the explanation of astrophysical disk--jet magnetized systems.
We stress here, that to fully explain
this complex dynamics, inclusion of density inhomogeneities,
gravity as well as rotation is crucial as shown in
\citep{BS-flow};\\
\citep{SY-DJ,yso};\\
\citep{jet-photon}.

In the dynamical Unified D/RD model explored in \cite{KS_2TRD},
it was shown that flow/outflow acceleration (of the bulk as well as the
fraction component), and the magnetic-field amplification is directly
proportional to the initial turbulent kinetic and/or magnetic energy.
In present study we show, that the  initial preparation of our
(complex) system fully defines the final fate of the composite
multi-temperature relativistic binary objects -- for the same
hot fraction temperature  $H_0$, and fraction coefficient $\alpha \ll 1$,
we found that, for a range of Beltrami parameters (magnetofluid coupling)
and relativistic helicities, the  system exhibits the explosive formation
of either large-scale / strong flows (dominated by the hot fraction) or the
large-scale / strong magnetic fields (see Figs.2,4 and Figs.3,5
for these 2 extreme cases).

For both dynamical RD/D and quasi--equilibrium scenarios,
the generated/accelerated outflows are extremely strong when both
background fluids are magnetically dominant;
hot outflows are several orders stronger than
the degenerate ones (see Figures 4 and 5 of present paper, specifically Fig.4).

For both scenarios: the large--scale fields formation process
was found to be less sensitive to the fraction parameter
$\alpha \ll 1$ of hot fluid but more sensitive to its temperature.
For some regimes of parameters (magnetically and/or kinetically
dominant mixed states), the growing accelerated flows in both fluids remain
sub-Alfv\'enic –- a scenario leading to strong magnetic-field
formation that grows as well (the explosive character of the
large--scale fields formation is determined by the rate
of the ambient system magneto-fluid coupling (Beltrami
parameters, helicities)) –- this could explain the formation
of large--scale magnetic fields during the envelope
phase of star accretion/WD evolution/binary systems.
In case of a fully magnetically dominant ambient system (for both fluids)
for realistic physical parameters,
the major part of its energy is transformed into the fast
locally super--Alfv\'enic large--scale composite flow energy
(magnetofluid coupling) as observed in a variety of
relativistic astrophysical outflows
–- this result is entirely due to the two-temperature character
of the initial composite relativistic system.
Interestingly, for explosive scenario degenerate flow energy grows/decreases
together with the magnetic field energy but remains sub-Alfv\'enic
while the hot flow behaves inversely -- in specific regimes,
it becomes super-Alfv\'enic (with Alfv\'en Mach-number
$> 10^6$) constituting a dominant part of
the bulk/composite flow energy (see Fig.4 of present study
for explosive scenario).

Thus, the formation of large--scale flows / magnetic fields
is guaranteed for our composite 2-temperature relativistic
binary  systems whether it follows the quasi--equilibrium
evolution through the catastrophic transformation of energies
or the dynamic scenario through the Unified RD/D process.
Interestingly, the quasi--equilibrium and the
dynamic approaches give similar results for the generation of
macro--scale velocity (magnetic) fields vindicating both.

\section{Summary and Conclusions}

Applying the quasi--equilibrium analysis to explore the explosive/erruptive
events, we have studied the ``evolution'' of multi--temperature
composite (multi--fluid) plasma systems (MF) often met
in astrophysical conditions. The overall quasi--neutral
plasma is composed of a mobile classical
ion component with two relativistic electron species (the bulk
degenerate electron gas and a small contamination of hot electrons).
The electron dynamics for both components are
described by the appropriate relativistic fluid equations.
The initial state is labelled by the  invariants (fluid--helicities and energy)
in conjunction with the initial and boundary conditions.

For this MF system:
\begin{itemize}
\item we have found, analytically, the condition of the catastrophic
transformation of energies. Catastrophe results when an initial
Quadruple Beltrami state is reduced to a final lower
energy state. Resulting scenarios for one such case for
the macro--scale velocity-- and magnetic field generation are delineated.

\item The most important qualitative result is that well defined
initial conditions can lead to a magnetically (kinetically) rich final
state -- in fact, in the former (latter) all of the flow
(magnetic) field energy is converted to the magnetic (velocity field)
energy at the catastrophe.

\item We show that the total flow energy
is dominated by the energy carried by the hot-fraction of the fluid.
\end{itemize}

This investigation laid down a framework -- a methodology that can
advance our understanding of the evolution of accreting astrophysical
objects/binaries. The dynamics/evolution is controlled jointly by
plasma flows and the magnetic field. It is shown, for example, that
macro--scale fast flow/outflow as well as primary macro--scale magnetic
fields could be generated from an appropriate mix of initial magnetic
and kinetic energy. The initial energy mix could be entirely short--scale.

The final state emerges as a consequence of a catastrophic transformation
of energies; we have shown that this transformation is guaranteed in
multi-temperature, multi-component systems as an intrinsic tendency
of flow acceleration/magnetic field amplification due to what
can be broadly labelled as magneto-fluid coupling.

The evolution physics has two distinct phases -- the first phase is on
a slow time scale but still can predict when a catastrophe might take place
by analyzing slowly evolving quasi equilibria that may change because of
slow changes in the surrounding environment. The catastrophe occurs when
these slow changes drive the system to a range of parameters that can,
no longer, sustain the said equilibrium.

However the very fast evolution in the vicinity (in time) of a catastrophe
requires a careful and proper time dependent treatment like the
dynamical Unified Reverse Dynamo/Dynamo mechanism \cite{msms};\\
\cite{KS_2TRD} or some other fast dynamics model. Naturally,
such a full treatment will be sensitive to many effects like
gravity, density/temperature inhomogeneities, and rotation.

However, what is fascinating is that slow evolving equilibrium
approach can predict whether a given initial configuration will
undergo a catastrophic transformation; it can also tell us how
different the transformed state is from the initial state!


\section{Acknowledgements}

Present work was partially supported by Shota Rustaveli
Georgian National Foundation Grant Project No. FR-22- 8273.
SMM's research is supported by  U.S. DOE under Grant Nos. DE-
FG02-04ER54742 and DE-AC02-09CH11466.

\end{document}